# Using Limited Trial Evidence to Credibly Choose Treatment Dosage when Efficacy and Adverse Effects Weakly Increase with Dose

Charles F Manski

Department of Economics and Institute for Policy Research, Northwestern University

May 2023

## Abstract

In medical treatment and elsewhere, it has become standard to base treatment intensity (dosage) on evidence in randomized trials. Yet it has been rare to study how outcomes vary with dosage. In trials to obtain drug approval, the norm has been to specify some dose of a new drug and compare it with an established therapy or placebo. Design-based trial analysis views each trial arm as qualitatively different, but it may be highly credible to assume that efficacy and adverse effects (AEs) weakly increase with dose. Optimization of patient care requires joint attention to both, as well as to treatment cost. This paper develops methodology to credibly use limited trial evidence to choose dosage when efficacy and AEs weakly increase with dose. I suppose that dosage is an integer choice $t \in (0, 1, \ldots, T)$, $T$ being a specified maximum dose. I study dosage choice when trial evidence on outcomes is available for only $K$ dose levels, where $K < T + 1$. Then the population distribution of dose response is partially rather than point identified. The identification region is a convex polygon determined by linear equalities and inequalities. I characterize clinical and public-health decision making using the minimax-regret criterion. A simple analytical solution exists when $T = 2$ and computation is tractable when $T$ is larger.

1. Introduction

1.1. Background

The need to choose a treatment intensity (dosage) arises in many settings. In medicine and elsewhere, it has become standard to base dosage on evidence obtained in randomized trials. Yet trial evidence has commonly been limited to comparison of at most a few dose levels.

Evidence on response to dosage has been particularly limited in the Phase III trials performed to obtain U.S. Food and Drug Administration (FDA) approval for marketing new drugs. These trials rarely study how patient outcomes vary with dosage. Instead, they typically specify some dose of a new drug and compare it with an established therapy or placebo. When evaluating drugs for treatment of cancer, the studied dose is often the *maximum tolerated dose*, a concept defined by National Cancer Institute (2023) as "the highest dose with acceptable side effects." To determine the maximum tolerated dose, pharmaceutical firms may compare several dose levels of a new drug in "dose-finding trials" undertaken as part of preliminary Phase II studies used to guide the design of Phase III trials (Viele *et al.*, 2015). However, Phase II sample sizes in are usually small and the findings not published.

*Illustration*: Clinicians seeking to achieve optimal adjuvant care for various cancers may choose between immunotherapy and surveillance. Immunotherapy aims to lower the risk of disease recurrence, but it may have toxicities that generate immune-related adverse effects (AEs), lowering quality of life or even yielding fatalities. It is credible to expect that, as dosage increases, the risk of recurrence falls and the risk of AEs rises. The reason is that the stimulation of the immune system produced by immunotherapy may destroy both malignant and healthy cells. Increasing dosage is also more costly. The optimal dosage appropriately weighs these opposing effects, which may be patient specific. The standard practice in Phase III studies has been to perform a two-armed trial comparing a specified dose of a new immunotherapy with an existing therapy or a placebo. Examples include recent trials that evaluated new immunotherapies for resectable melanoma (Eggermont *et al*., 2015, Weber *et al*. 2017, Eggermont et al., 2018). These trials provide no direct information about outcomes with alternative dosages that might yield better combinations of efficacy and AEs. Mathematical modeling of dose response suggest that the dose levels specified in these trials could be lowered by at least fifty percent without reduction of efficacy (Ratain and Goldstein, 2018; Peer *et al*., 2022). □

I wrote above that Phase III studies provide no "direct" information about alternative dosages. The qualifier "direct" stems from the fact that the prevalent *design-based* analysis of randomized trials views each trial arm as qualitatively different. Consider, for example, a three-arm trial comparing treatments (A, B, C). In design-based analysis, findings with arm A are not used to draw conclusions about arms B and C, and vice versa.

Design-based analysis is well-grounded when treatment arms differ qualitatively from one another. However, this commonly is not so when comparing dosages of treatments. If A, B, and C are increasing dosages of an immunotherapy or another cancer treatment such as chemotherapy or radiation, there is good reason to think that efficacy and AEs both increase with dosage. Monotonicity of efficacy and AEs is similarly credible with drug treatment of diseases other than cancer. Thus, the findings for arm A provide indirect information about arms B and C, in the form of bounds on outcomes. Designed-based analysis does not use this information.

Beyond drug treatment, it similarly may be credible to assume that efficacy and AEs increase with dosage in other aspects of medicine and other realms of treatment. In medicine, the efficacy of radiological scans in detecting disease increases with dosage but so do the AEs generated by radiation. In agriculture, increasing the dosage of herbicides and pesticides may increase crop yields but produce greater environmental AEs and is more costly. In criminal justice, efficacy in preventing crime may increase with the intensity of proactive policing activities such as stop and frisk, but increasing these activities may harm the community by infringing on the privacy of innocent persons (Manski and Nagin, 2017).

In medical research, a shortcoming of prevalent trial analysis of dosage has been the practice of studying efficacy and AEs separately, viewing the former as the primary outcome and the latter as a secondary outcome. This is reasonable if efficacy has a much larger effect on patient health than AEs. However, this asymmetry may not prevail when considering dosage for treatment of cancers and some other diseases. Optimization of patient care requires joint attention to efficacy and AEs. This can be achieved by specifying and studying a composite measure of patient health that varies appropriately with both outcomes.

It is concerning that medical literature makes frequent references to optimal patient care, without clearly defining optimality. Consider, for example, a new Food and Drug Administration (2023) draft Guidance for Industry with the title "Optimizing the Dosage of Human Prescription Drugs and Biological Products for the Treatment of Oncologic Diseases." The document does not clearly explain what it means to optimize dosage. It states (page 1, footnote 2): "Optimal dosage is the dosage that can maximize the benefit/risk profile or provide the desired therapeutic effect while minimizing toxicity." This brief attempt to define optimal dosage does not recognize that efficacy and toxicity both increase with dosage, in a manner that may be patient specific. The tension between these forces makes it impossible to find a dosage that simultaneously maximizes therapeutic effect while minimizing toxicity. The optimal dosage must appropriately weigh these opposing forces.

## 1.2. This Paper

The long-run solution to the current dearth of trial evidence on dose response should be to design and execute new trials that enrich the available data. For now, it is important to use the existing data effectively. This goal motivates the present paper, which develops methodology to credibly use limited trial evidence to choose treatment dosage when efficacy and AEs weakly increase with dose. I bring to bear principles of partial identification analysis and minimax-regret (MMR) decision making described in Manski (2018, 2019).

I suppose that dosage is an integer choice variable t ∈ (0, 1, . . . , T) among feasible dose levels, where T is a specified maximum dose. I assume that the objective is to maximize expected patient welfare, which is a function of treatment efficacy and AEs, minus expected treatment cost. To simplify the analysis, I consider illness and adverse effects to be binary events (yes/no), rather than outcomes that vary in severity. I also suppose that dosage is a one-time choice rather than a dynamic decision.

The central difficulty addressed in the paper is that trial evidence on outcomes is available for only K dose levels, where K < T + 1. Assuming only that efficacy and AEs are weakly increasing in dosage, the population distribution of dose response is partially rather than point identified. I find that the identification region is a convex polygon determined by linear equalities and inequalities.

The identification problem makes it logically infeasible to choose dosage to maximize expected welfare. Nevertheless, a clinician or public-health planner can make a reasonable dosage choice under ambiguity. Whereas a clinician chooses a treatment for a single patient, a planner allocates a population of patients across dose options. I study decision making using the minimax-regret criterion. I find that a simple analytical solution exists when T = 2 and computation is tractable when T is larger.

Section 2 sets up the dosage-choice problem. Section 3 introduces the central maintained assumption that efficacy and AEs weakly increase with dosage. I study identification of the distribution of dose response under this assumption, considering the sample size in the trial to be large enough to make sampling imprecision negligible. Section 4 analyzes dosage choice in this setting, with brief consideration of decision making when sampling imprecision is not negligible. Section 5 discusses extensions of the analysis that warrant future research.

The analysis of identification in this paper differs in several respects from the study of monotone treatment response in Manski (1997). That article analyzed partial identification using observational data when a univariate outcome varies monotonically with treatment intensity. It did not study decision making. The present identification setting differs because the (efficacy, AE) outcome is bivariate, with increasing dosage improving the former outcome but worsening the latter one. It also differs because trial data are more informative than observational data. A further difference is that outcomes are binary here, whereas they were only required to be bounded in the earlier paper.

The present work differs substantially from the large body of research that assumes a parametric dose response model of a univariate outcome in order to point-identify and estimate dose response. A leading case in pharmacology is nonlinear least squares estimation of a certain three-parameter version of the *Hill equation* predicting a univariate biological response generated by a scalar drug concentration. See Gesztelyi *et al.* (2012), equation 8. I am not aware of studies that specify and estimate a parametric model of the bivariate (efficacy, AE) outcome. The norm has been to study efficacy and AEs separately.

It may appear that point estimation of a parametric model of the bivariate outcome would be valuable to patient care, as one might then aim to approximately optimize dose choice using the point estimate. The value is real if the specified parametric dose response model is known to be correct, but it is illusory otherwise. Concern with the credibility of such a model motivates the present research on dosage choice under ambiguity. The decision criterion used to formalize choice under ambiguity is consequential. Section 4 gives several reasons that motivate focus on MMR rather than alternatives such as subjective Bayes and maximin decisions.

The analysis in this paper has potential application to diverse non-medical problems of dosage choice in settings with monotone response of (efficacy, AEs) to dosage. For concreteness, I use medical language throughout, considering a clinician who treats a patient or a public health planner that treats a population. Within the medical realm, I sometimes specifically refer to treatment with drugs, this being a highly important area of application.

## 2. The Dosage-Choice Problem

Letting j label a patient, I suppose that two dose-dependent treatment outcomes $[d_j(t), e_j(t)]$ determine the patient's health, or more generally, welfare. Given dose t, patient j may experience a disease $[d_j(t) = 1$ if yes, $d_j(t) = 0$ if no] and/or an AE $[e_j(t) = 1$ if yes, $e_j(t) = 0$ if no]. Thus, the bivariate outcome $[d_j(t), e_j(t)]$ takes one of the four values [(0, 0), (1, 0), (0, 1), (1, 1)]. The simplifying assumption that disease and AEs, are binary outcomes rather than more complex phenomena enables transparent analysis of the tension between efficacy and AEs that arises in dosage choice.

Welfare is a function of these outcomes, denoted $w_j[d_j(t), e_j(t)]$. Treatment cost, including the monetary and non-monetary costs of administering treatment, is a function $g_j(t)$ of the dose level, with cost measured in the same units as welfare. In medical contexts, welfare is often measured as quality-adjusted life years (QALYS); see Loomis and McKenzie (1989).

I index outcomes, welfare, and cost by j, permitting full heterogeneity across the population. It is realistic to suppose that disease and AEs both lower welfare; thus, $w_j(0, 0) > \max [w_j(0, 1), w_j(1, 0)] \geq \min [w_j(0, 1), w_j(1, 0)] > w_j(1, 1)$ for all patients j. It is also realistic to suppose that treatment cost is non-negative and weakly increases with dosage. My analysis does not require these conditions, but I use them when reporting illustrative computational findings.

I assume that the objective is to maximize welfare net of cost. Interpretation of optimal dosage depends on the knowledge that a clinician of other planner possesses. A highly idealized interpretation supposes that the planner has perfect foresight, knowing $w_j[d_j(t), e_j(t)]$ and $g_j(t)$ for all $t \in \{0, 1, 2, \ldots, T\}$. Then an optimal dose solves the problem max $_{t = 0, \ldots, T}$ $w_j[d_j(t), e_j(t)] - g_j(t)$.

Perfect foresight is too unrealistic an assumption to provide a useful benchmark for decision making. It has been common in medical economics to study optimal care supposing that a clinician has objectively correct probabilistic expectations; see, for example, Phelps and Mushlin (1988). Assume that the clinician observes certain patient attributes, such as age, sex, and health history. Viewing j as a member of a population J of patients who have the same observable attributes, assume that the clinician knows the dose-dependent distributions $p\{w[d(t), e(t)], g(t)\}$, $t \in \{0, 1, 2, \ldots, T\}$ of (welfare, cost) over J and chooses a dose that maximizes expected net welfare. Then an optimal dose solves max $_{t = 0, \ldots, T}$ $E\{w[d(t), e(t)]\} - E[g(t)]$.

The above knowledge scenario is still typically unrealistic. The standard protocol for data collection in randomized clinical trials yields empirical evidence only on patient disease and adverse-effect outcomes, not on welfare as a function of these outcomes. Medical economists conduct separate studies that sample patients and aim to learn their welfare functions, by questioning them about the choices they would make if they were to experience hypothetical disease and AEs; e.g., Basu and Meltzer (2007) and Devlin and Brooks (2017). However, the patients in these studies typically are different persons than the subjects in clinical trials.

Estimation of dose-dependent expected welfare may be feasible if it is credible to assume that welfare functions are mean-independent of disease and adverse-effect outcomes; that is, $E[w(h, i)|d(t) = h, e(t) = i] = E[w(h, i)]$, $h = 0, 1$ and $i = 0, 1$. Then expected welfare with dose t is

(1)  $E\{w[d(t), e(t)]\} = E[w(0, 0)]\cdot p[d(t) = 0, e(t) = 0] + E[w(1, 0)]\cdot p[d(t) = 1, e(t) = 0]$

$\qquad\qquad\qquad + E[w(0, 1)]\cdot p[d(t) = 0, e(t) = 1] + E[w(1, 1)]\cdot p[d(t) = 1, e(t) = 1] - E[g(t)].$

A study of patient choice in hypothetical scenarios enables estimation of the outcome-specific expected welfare function $E[w(\cdot, \cdot)] = \{E[w(1, 1)], E[w(1, 0)], E[w(0, 1)], E[w(0, 0)]\}$. A randomized trial with $T + 1$ arms $t \in \{0, 1, 2, \ldots, T\}$ enables estimation of the dose-dependent outcome distributions $p[d(t), e(t)]$, $t \in \{0, 1, 2, \ldots, T\}$. The prevalent practice in articles presenting trial findings has been to report only marginal outcome distributions $p[d(t)]$ and $p[e(t)]$, not bivariate distributions $p[d(t), e(t)]$. I will ignore this inadequacy of the literature, as it can be remedied with access to subject-specific trial data.

I henceforth assume that (1) holds and that research has revealed $E[w(\cdot, \cdot)]$. I also assume that mean treatment cost $E[g(t)]$ is known. My concern is that it has been rare to perform the aforementioned $(T + 1)$-armed trial. Instead, for some integer K with $1 < K < T + 1$, the standard practice has been to perform a K-armed trial, assigning subjects to K specified dosages, $t_k$, $k = 1, \ldots, K$. With this design, a trial directly enables one to estimate only the outcome distributions $p[d(t_k), e(t_k)]$, $k = 1, \ldots, K$, not $p[d(t), e(t)]$ for other dosages. Supposing that the trial is large enough that sampling imprecision is negligible, design-based trial analysis enables solution of the dose-constrained optimization problem $\max_{k=1,\ldots,K} E\{w[d(t_k), e(t_k)]\} - E[g(t_k)]$, but not the unconstrained problem $\max_{t=0,\ldots,T} E\{w[d(t), e(t)]\} - E[g(t)]$.

The incompleteness of design-based analysis may be only a minor concern when the K treatment arms $(t_k, k = 1, \ldots, K)$ well-approximate the range $t \in \{0, 1, 2, \ldots, T\}$ of feasible dose levels. It can become a major difficulty when K is small relative to $T + 1$. Particularly incomplete are standard Phase III drug trials where K = 2, with $t_1 = 0$ and $t_2$ being a specified positive dose level. Then it becomes important to explore credible extrapolation beyond the trial design, enabling informative conclusions to be drawn about outcomes with dose levels that are not in the design. This is the subject addressed below.

## 3. Identification with Monotone Dose Response

### 3.1. Monotone Dose Response

As discussed in the Introduction, there often is good reason to expect that efficacy and AEs increase with dosage. Disease being a binary outcome, a patient who remains free of disease with a low dose would remain free of disease with a higher dose. AEs being a binary outcome, a patient who has an adverse effect with a low dose would have one with a higher dose. Formally, I pose these assumptions of monotone dose response:

*Monotone Dose Response*: Consider dose levels (s, t) with $t > s$. For each patient $j \in J$,

Monotone Efficacy (ME): $d_j(s) = 0 \Rightarrow d_j(t) = 0$.

Monotone AEs (MT): $e_j(s) = 1 \Rightarrow e_j(t) = 1$.

A less concise but analytically helpful way to express monotone dose response uses the concept of patient-specific threshold dose levels.

*Monotone Dose Response Expressed Through Threshold Dose Levels*: For each patient j ∈ J,

Monotone Efficacy (ME): Either $d_j(t) = 1$, all $t \in \{0, 1, 2, \ldots, T\}$, or there exists a threshold dose $t_{dj} \in \{0, 1, 2, \ldots, T\}$ such that $d_j(t) = 1$ for $t < t_{dj}$ and $d_j(t) = 0$ for $t \geq t_{dj}$. To concisely express when $d_j(t) = 1$, all $t \in \{0, 1, 2, \ldots, T\}$, it suffices to define an infeasible threshold dose $t_{dj} = T + 1$.

Monotone AEs (MT): Either $e_j(t) = 0$, all $t \in \{0, 1, 2, \ldots, T\}$ or there exists a threshold dose $t_{ej} \in \{0, 1, 2, \ldots, T\}$ such that $e_j(t) = 0$ for $t < t_{ej}$ and $e_j(t) = 1$ for $t \geq t_{ej}$. To concisely express when $e_j(t) = 0$, all $t \in \{0, 1, 2, \ldots, T\}$, it suffices to define an infeasible threshold dose $e_{dj} = T + 1$.

The two versions of (ME)−(MT) are equivalent, as $d_j(t) = 1[t < t_{dj}]$ and $e_j(t) = 1[t \geq t_{ej}]$.

## 3.2. Identification with Limited Trial Evidence

I consider identification with limited trial evidence. Let $q(t_d, t_e)$ denote the population distribution of threshold dose levels. In the absence of trial evidence, $q(t_d, t_e)$ may be any bivariate distribution on $\{0, \ldots, T + 1\} \times \{0, \ldots, T + 1\}$. Thus, there are $(T + 2)^2$ component probabilities. The non-negativity of probabilities and the Law of Total Probability imply that $q(t_d, t_e)$ satisfies these basic conditions:

(2)  $q(t_d = h, t_e = i) \geq 0, h \in \{0, 1, 2, \ldots, T + 1\}, i \in \{0, 1, 2, \ldots, T + 1\}.$

(3)  $1 = \displaystyle\sum_{h = 0}^{T + 1} \sum_{i = 0}^{T + 1} q(t_d = h, t_e = i).$

The outcome distributions $p[d(t), e(t)]$, $t \in \{0, 1, 2, \ldots, T\}$ have $4(T + 1)$ component probabilities. Equations (4a)−(4d) express the outcome probabilities in terms of $q(t_d, t_e)$. For $t \in \{0, 1, 2, \ldots, T\}$,

(4a)  $p[d(t) = 0, e(t) = 0] = q(t_d \leq t, t_e > t) = \sum\limits_{h=0}^{t} \sum\limits_{i=t+1}^{T+1} q(t_d = h, t_e = i).$

(4b)  $p[d(t) = 1, e(t) = 0] = q(t_d > t, t_e > t) = \sum\limits_{h=t+1}^{T+1} \sum\limits_{i=t+1}^{T+1} q(t_d = h, t_e = i).$

(4c)  $p[d(t) = 0, e(t) = 1] = q(t_d \leq t, t_e \leq t) = \sum\limits_{h=0}^{t} \sum\limits_{i=0}^{t} q(t_d = h, t_e = i).$

(4d)  $p[d(t) = 1, e(t) = 1] = q(t_d > t, t_e \leq t) = \sum\limits_{h=t+1}^{T+1} \sum\limits_{i=0}^{t} q(t_d = h, t_e = i).$

Given limited trial evidence, the above preliminaries immediately yield the identification regions for $q(t_d, t_e)$, $p[d(t), e(t)]$, and $E\{w[d(t), e(t)]\}$. Proposition 1 gives the findings. I denote the identification region for $q(t_d, t_e)$ as Q. The findings for identification of $p[d(t), e(t)]$ and $E\{w[d(t), e(t)]\}$, $t \in \{0, 1, 2, \ldots, T\}$ follow immediately from determination of Q. The identification regions for $p[d(t), e(t)]$, $t \in \{0, 1, 2, \ldots, T\}$ and $E\{w[d(t), e(t)]\}$, $t \in \{0, 1, 2, \ldots, T\}$ are denoted P and W.

*Proposition 1*: Let (ME) and (MT) hold. Let a K-armed trial assigning subjects to dosages, $t_k$, $k = 1, \ldots, K$ reveal the outcome distributions $p[d(t_k), e(t_k)]$, $k = 1, \ldots, K$. Then Q comprises all threshold-dose distributions $q(t_d, t_e)$ that satisfy (2), (3), and, for $k = 1, \ldots, K$,

(5a)  $p[d(t_k) = 0, e(t_k) = 0] = q(t_d \leq t_k, t_e > t_k) = \sum_{h=0}^{t_k} \sum_{i=t_k+1}^{T+1} q(t_d = h, t_e = i),$

(5b)  $p[d(t_k) = 1, e(t_k) = 0] = q(t_d > t_k, t_e > t_k) = \sum_{h=t_k+1}^{T+1} \sum_{i=t_k+1}^{T+1} q(t_d = h, t_e = i),$

(5c)  $p[d(t_k) = 0, e(t_k) = 1] = q(t_d \leq t_k, t_e \leq t_k) = \sum_{h=0}^{t_k} \sum_{i=0}^{t_k} q(t_d = h, t_e = i),$

(5d)  $p[d(t_k) = 1, e(t_k) = 1] = q(t_d > t_k, t_e \leq t_k) = \sum_{h=t_k+1}^{T+1} \sum_{i=0}^{t_k} q(t_d = h, t_e = i).$

P comprises all collections of distributions $p[d(t), e(t)]$, $t \in \{0, 1, 2, \ldots, T\}$, $t \in \{0, 1, 2, \ldots, T\}$ that satisfy (4a)−(4d) for some element of Q. W comprises all vectors of means $E\{w[d(t), e(t)]\}$, $t \in \{0, 1, 2, \ldots, T\}$ that satisfy (1) for some element of P.  □

Proposition 1 shows that $q(t_d, t_e)$ is partially identified. Recall that $q(t_d, t_e)$ has $(T + 2)^2$ component probabilities. These solve one linear equation by The Law of Total Probability (3) and 4K more by (5a)−(5d). For each k, the Law of Total Probability implies that only three of the four equations are linearly independent. Hence, $q(t_d, t_e)$ solves $3K + 1$ non-redundant linear equations in $(T + 2)^2$ unknowns, plus the linear inequalities (2), making Q a $(T + 2)^2 - (3K + 1)$ dimensional convex polygon. $q(t_d, t_e)$ is partially identified even when the trial design includes all feasible dose levels. Then $K = T + 1$ and $q(t_d, t_e)$ solves $3T + 4$ linear equations, which is less than $(T + 2)^2$ for all $T \geq 1$.

The trial evidence point-identify $p[d(t_k), e(t_k)]$, $k = 1, \ldots, K$. The problem of concern is identification of outcome distributions at dose levels t not included in the trial design. $p[d(t), e(t)]$ is a linear function of $q(t_d, t_e)$ and Q is a convex polygon. Hence, P is a convex polygon.

3.3. Restricting the Distribution of Threshold Doses

The above analysis of identification assumes only monotone dose response, without restricting the distribution $q(t_d, t_e)$ of threshold doses. Assumptions on this distribution may have identifying power and be credible in some applications. I will briefly discuss some possibilities here.

*No AEs Without Treatment*: A highly credible assumption is that AEs cannot occur when the dose level is zero. Formally, assume that $q(t_d = h, t_e = 0) = 0$, $h \in \{0, 1, 2, \ldots, T + 1\}$. The assumption fixes $T + 2$ components of $q(t_d, t_e)$, reducing the number of unknown components from $(T + 2)^2$ to $(T + 2)(T + 1)$. It implies that $p[d(0) = 0, e(0) = 1] = p[d(0) = 1, e(0) = 1] = 0$.

*Concurrent Thresholds for Efficacy and AEs*: In some contexts, it is credible to pose assumptions on the cross-patient association between thresholds $t_d$ and $t_e$. A positive association is credible in the treatment of cancer with immunotherapy because stimulation of the immune system increases with dose level, making both tumor reduction and AEs more likely. Patient immune systems vary genetically and epigenetically, so the degree to which immunotherapy generates immune responses may vary across patients.

There are many ways to formalize positive association as an assumption on $q(t_d, t_e)$. A polar case assumes that each patient has one threshold dose that jointly prevents disease recurrence and generates an AE. That is, there exists a patient-specific dose $t_{*j}$ such that $t_{*j} = t_{dj} = t_{ej}$, all j. Then $q(t_d = h, t_e = i) = 0$ whenever $h \neq i$, reducing the number of unknown components of $q(t_d, t_e)$ from $(T + 2)^2$ to $T + 2$. The assumption implies that, for each dose level t, $p[d(t) = 0, e(t) = 0] = p[d(t) = 1, e(t) = 1] = 0$.

*Statistical Independence of Efficacy and AE Thresholds*: Whereas the biological mechanism of immunotherapy suggests positive association between thresholds, in other treatment contexts separate physiological processes may make a treatment efficacious and yield AEs. In these cases, it may be reasonable to assume statistical independence of $t_d$ and $t_e$. Then $q(t_d = h, t_e = i) = q(t_d = h) \cdot q(t_e = i)$ for all $(h, i)$, reducing the number of unknown components of $q(t_d, t_e)$ from $(T + 2)^2$ to $2(T + 2)$. The assumption implies that, for each dose level t, $p[d(t), e(t)] = p[d(t)] \cdot p[e(t)]$.

## 4. Dosage Choice with Monotone Dose Response

I now consider several versions of the dosage choice problem. As in Section 3, Sections 4.1 and 4.2 assume that a K-armed trial assigning subjects to dosages, $t_k$, k = 1, . . . , K reveal $p[d(t_k), e(t_k)]$, k = 1, . . . , K. Section 4.3 briefly supposes that only sample data are observed.

I consider decision making from the perspectives of a clinician treating one patient and a public-health planner treating a population with the same distribution of dose thresholds as J. The clinician must choose a single treatment t for the patient and is not permitted to randomize the choice. The planner can choose a fractional treatment allocation $\delta(t)$, t ∈ {0, 1, 2, . . . , T} in the unit simplex $\Delta$ on $R^{T+1}$. Thus, the planner has a richer set of treatment possibilities than does the clinician.

The central problem for the decision maker is incomplete knowledge of the dose threshold distribution $q(t_d, t_e)$. The familiar Bayesian prescription for coping with uncertainty is to assert a subjective probability distribution on unknown quantities and maximize subjective expected welfare. Objective expected welfare $E\{w[d(t), e(t)]\}$ is linear in $q(t_d, t_e)$. Hence, a Bayesian decision maker would assert a subjective distribution, say $\pi$, on the space of all possible threshold dose distributions, replace the unknown $q(t_d, t_e)$ with its subjective mean $\int q(t_d, t_e)d\pi$, and choose the dose that would be optimal if $\int q(t_d, t_e)d\pi$ were the actual threshold dose distribution.

The Bayes decision criterion is attractively simple, but it presumes that the decision maker is able to place a credible subjective distribution on $q(t_d, t_e)$. Bayesians have long struggled to provide guidance on specification of subjective distributions and the matter continues to be controversial. When one finds it difficult to assert a credible subjective distribution, Bayesians may suggest use of some default subjective distribution, called a "reference" or "conventional" or "objective" prior; see Berger (2006). However, there is no consensus on the prior that should play this role. The chosen prior critically affects decisions.

Inability to credibly specify a prior motivates study of criteria for decision making under ambiguity, which do not place subjective distributions on unknown quantities. Two prominent approaches are maximin and minimax regret (MMR). I evaluate decisions by their maximum regret. Section 4.1 explains this focus.

### 4.1. Minimax-Regret Decisions

I first pose the MMR criterion abstractly and then apply it to dosage choice. Consider a planner who faces a choice set C and believes that the true state of nature lies in a specified state space S. An objective function $f(\cdot, \cdot)$: $C \times S \to R^1$ maps actions and states into welfare. The planner wants to maximize true welfare, but he does not know the true state. In settings without sample data, the minimax-regret (MMR) criterion solves the problem

$$(6) \quad \min_{c \in C} \ \max_{s \in S} \ [\max_{d \in C} f(d, s) - f(c, s)].$$

Here $\max_{d \in C} w(d, s) - w(c, s)$ is the *regret* of action c in state s; that is, the degree of suboptimality.

In the present setting, the state space indexes the feasible distributions of threshold doses derived in Proposition 1; thus, S = Q. In the clinical case, the choice set comprises all feasible dose levels; thus, C = {0, 1, 2, . . . , T}. In the public-health case, the choice set comprises all fractional allocations; thus, C = Δ.

In the clinical case, the objective function in state s is $f[t, q_s(t_d, t_e)] = E_s\{w[d(t), e(t)]\}$, where $q_s(t_d, t_e)$ is the distribution indexed by state s and $E_s$ denotes expectation with respect to this distribution. In the public-health case, the objective function is the mean of the expected welfare function with respect to the treatment allocation; thus, $f[\delta(\cdot), q_s(t_d, t_e)] = \sum_t \delta(t) \cdot E_s\{w[d(t), e(t)]\}$.

I study dosage choice using the MMR criterion because, as discussed elsewhere (e.g., Manski, 2018, 2019, 2021), this criterion has attractive properties when decision makers must make choices under ambiguity. MMR should not be confused with maximin, which solves the problem $\max_{c \in C} \min_{s \in S} f(c, s)$.

The two differ except in special cases, particularly when optimal welfare is invariant across states. Whereas maximin considers only the worst outcome that an action may yield across states, MMR considers the worst outcome relative to what is achievable in a given state. Thus, MMR does not express the ultra-pessimism characteristic of the maximin criterion.

A conceptual appeal of using maximum regret to measure performance is that it quantifies how lack of knowledge of the true state of nature diminishes the quality of decisions. The term "maximum regret" is a shorthand for the maximum sub-optimality of a decision rule across feasible states of nature. A decision rule with small maximum regret is uniformly near-optimal across all states. This is a desirable property.

## 4.2. Computation of MMR Decisions

Exact computation of the MMR solution is tractable in the clinical case when T is not extremely large. An alternative expression of (6) reverses the two max operations to obtain

$$(6') \qquad \min_{c \in C} \quad \max_{d \in C} \quad \max_{s \in S} \, [f(d, s) - f(c, s)].$$

Holding c and d fixed, consider the inner maximization over s. When c = d, the maximum is 0 as $f(d, s)$ = $f(c, s)$ for all values of s. When c $\neq$ d, $f(d, s) - f(c, s)$ is a linear function of $q_s(\cdot, \cdot)$ in the setting of this paper. The feasible distributions Q satisfy the linear inequalities (2) and equations (3) and (5a)−(5d). Hence, $\max_{s \in S} [f(d, s) - f(c, s)]$ is a linear programming problem, which can be solved exactly using standard algorithms. Considering all c $\neq$ d, determination of the MMR dosage requires solution of $T \cdot (T + 1)$ linear programming problems. This is tractable when T is not extremely large.

In the public-health case, $\max_{s \in S} [f(d, s) - f(c, s)]$ remains a linear programming problem, but $f(d, s) - f(c, s)$ is a different linear function of $q_s(\cdot, \cdot)$. The set of feasible dosage allocations is the entire simplex $\Delta$ on $R^{T+1}$ rather than only the vertices, each of which places probability one on a single treatment. When T > 2, this makes exact computation of the MMR dosage allocation infeasible. However, it is feasible to numerically approximate the solution by using a suitable finite grid to approximate $\Delta$ and then solving the associated finite number of linear programming problems.

### 4.2.1. Public-Health Dosage Allocation when T = 2

Exact computation of public-health MMR dosage allocation is feasible, indeed simple, when $T = 2$, $K = 2$, $t_1 = 0$, and $t_2 = 2$. Thus, there is no trial evidence for $t = 1$. This special case is important in dosage choice for cancer drugs. As discussed in the Introduction, Phase III drug trials commonly compare zero dose and a maximum tolerable dose, without evaluating a smaller positive dose.

In this case, welfare net of treatment cost is point identified for $t = 0$ and $t = 2$. Let them be denoted $\omega_0 \equiv E\{w[d(0), e(0)]\} - E[g(0)]$ and $\omega_2 \equiv E\{w[d(2), e(2)]\} - E[g(2)]$. Welfare net of treatment cost is partially identified for $t = 1$, being known to lie in a computable interval, say $[\omega_{1L}, \omega_{1U}]$, determined by the identification analysis of Section 3. There is no ambiguity if $\omega_{1L} \geq \max(\omega_0, \omega_2)$ or $\omega_{1U} \leq \max(\omega_0, \omega_2)$. I focus on settings with $\omega_{1L} < \max(\omega_0, \omega_2) < \omega_{1U}$, where there is ambiguity.

Manski (2009) studied MMR allocation of a population between two undominated treatments. The analysis applies here, where dose $t = 1$ and either $t = 0$ or $t = 2$ are undominated. Manski (2009), equation (8) shows that the MMR allocation assigns positive fractions of patients to $t = 1$ and to the other dose with the higher value of $\omega$. If $\omega_2 > \omega_0$, dose $t = 0$ is dominated and so receives zero allocation. The fraction of the population assigned to $t = 1$ is $(\omega_{1U} - \omega_2)/(\omega_{1U} - \omega_{1L})$ and the fraction assigned to $t = 2$ is $(\omega_2 - \omega_{1L})/(\omega_{1U} - \omega_{1L})$. Symmetrically, if $\omega_2 < \omega_0$, $t = 2$ is dominated and receives zero allocation. The fraction assigned to $t = 1$ is $(\omega_{1U} - \omega_0)/(\omega_{1U} - \omega_{1L})$ and the fraction assigned to $t = 0$ is $(\omega_0 - \omega_{1L})/(\omega_{1U} - \omega_{1L})$. If $\omega_2$ and $\omega_0$ both equal some value $\omega$, the fraction assigned to $t = 1$ is $(\omega_{1U} - \omega)/(\omega_{1U} - \omega_{1L})$, with the remainder assigned in any proportions to $t = 0$ and $t = 2$.

The finding that the MMR criterion assigns a positive fraction of patients to $t = 1$ demonstrates the difference between MMR and maximin decision making. When $\omega_{1L} < \max(\omega_0, \omega_2) < \omega_{1U}$, a maximin planner does not allocate anyone to $t = 1$. A planner using the MMR criterion does so.

### 4.2.2. Numerical Illustrations

I use a specific case of the design with $T = 2$, $K = 2$, $t_1 = 0$, and $t_2 = 2$ to illustrate numerically. Let expected patient welfare with each $(d, e)$ outcome be as follows:

$E[w(0, 0)] = 1$, $E[w(1, 0)] = 0.25$, $E[w(0, 1)] = 0.75$, $E[w(1, 1)] = 0$.

This quantifies the usual situation where it is best to experience neither disease nor an adverse event, and worst to experience both. Experiencing one or the other is intermediate, disease being more harmful on average than an adverse event.

Let the threshold dose distribution be as follows:

$q(t_d = h, t_e = 0) = 0$, $h \in \{0, 1, 2, 3\}$;   $q(t_d = h, t_e = i) = 1/12$, $h \in \{0, 1, 2, 3\}$, $i \in \{1, 2, 3\}$.

The first part of the specification expresses the credible assumption that AEs cannot occur without treatment. The uniform distribution in the second part of the specification is only intended to be illustrative, as the actual distribution will be context specific.

The outcome probabilities implied by this threshold dose distribution are

$p[d(0) = 0, e(0) = 0] = 0.25$      $p[d(0) = 0, e(0) = 1] = 0$
$p[d(0) = 1, e(0) = 0] = 0.75$      $p[d(0) = 1, e(0) = 1] = 0.$

$p[d(1) = 0, e(1) = 0] = 0.333$      $p[d(1) = 0, e(1) = 1] = 0.167$
$p[d(1) = 1, e(1) = 0] = 0.333$      $p[d(1) = 1, e(1) = 1] = 0.167.$

$p[d(2) = 0, e(2) = 0] = 0.25$      $p[d(2) = 0, e(2) = 1] = 0.5$
$p[d(2) = 1, e(2) = 0] = 0.083$      $p[d(2) = 1, e(2) = 1] = 0.166.$

The values of expected welfare unconditional on $(d, e)$ are

$E\{w[d(0), e(0)]\} = 0.4375$,  $E\{w[d(1), e(1)]\} = 0.542$, $E\{w[d(2), e(2)]\} = 0.6458$.

The values for $p[d(0), e(0)]$, $p[d(2), e(2)]$, $E\{w[d(0), e(0)]\}$, and $E\{w[d(2), e(2)]\}$ are point-identified by the trial evidence. Application of Proposition 1 shows that the identification region for $p[d(1), e(1)]$ is a convex polygon in $R^4$, whose projections on the four axes are these bounds:

$p[d(1) = 0, e(1) = 0] \in [0 ,0.75]$      $p[d(1) = 0, e(1) = 1] \in [0, 0.5]$
$p[d(1) = 1, e(1) = 0] \in [0.083, 0.75]$      $p[d(1) = 1, e(1) = 1] \in [0, 0.67].$

The identification region for $E\{w[d(1), e(1)]\}$ is the interval $[0.2708, 0.8125]$.

Now consider MMR clinical and public-health dosage choice with various cost functions g(t). Here are some illustrative findings:

g(t) = 0, all t: The MMR clinical dose is t = 2, with MMR value 0.167. The public-health dose allocation is (0, 0.308, 0.692) to doses (0, 1, 2), with MMR value 0.116.

g(t) = (0.05)t: The MMR clinical dose is t = 2, with MMR value 0.217. The public-health dose allocation is (0, 0.4, 0.6) to doses (0, 1, 2), with MMR value 0.13.

g(t) = (0.1)t: The MMR clinical dose is t = 2, with MMR value 0.267. The public-health dose allocation is (0, 0.49, 0.51) to doses (0, 1, 2), with MMR value 0.136.

g(t) = (0.15)t: The MMR clinical dose is t = 0, with MMR value 0.225. The public-health dose allocation is (0.59, 0.41, 0) to doses (0, 1, 2), with MMR value 0.132.

g(0) = g(1) = 0, g(2) = 0.30: The MMR clinical dose is t = 1, with MMR value 0.167. The public-health dose allocation is (0.308, 0.692, 0) to doses (0, 1, 2), with MMR value 0.115.

Observe that the clinical decision is t = 2 or t = 0 with each of the four linear cost functions, but it is t = 1 with the nonlinear cost function. Thus, a clinician using the MMR criterion may choose a dose level excluded from the trial design.

## 4.3. Decisions with Finite-Sample Trial Data

Statistical decision problems suppose that the planner observes data generated by a sampling distribution, which varies with the state of nature. To express this, let $\Psi$ denote the sample space; that is, the set of samples that may be drawn. Let the possible sampling distributions be denoted $(Q_s, s \in S)$. The literature assumes that the sample space does not vary with s and is known, whereas the sampling distribution varies with s and is unknown. A statistical decision function, $c(\cdot): \Psi \to C$ maps the sample data into a chosen action.

An SDF is a deterministic function after realization of the sample data, but it is a random function ex ante. Hence, an SDF generically makes a randomized choice of an action. Statistical decision theory evaluates the performance of SDF $c(\cdot)$ in state s by $Q_s\{f[c(\psi), s]\}$, the ex-ante distribution of welfare across realizations $\psi$ of the sampling process. In abstraction, the statistical version of the MMR criterion is

(7)     $\min_{c(\cdot)\,\in\,\Gamma}\quad \max_{s\,\in\,S}\quad (\max_{d\,\in\,C} f(d,\,s) - E_s\{f[c(\psi),\,s]\}).$

In the present setting, the choice set and the objective function are as in Section 4.1. Sample data do not reveal the outcome distributions $p[d(t_k),\,e(t_k)]$, $k = 1, \ldots, K$; the data only enable estimation. Hence, the present state space is larger than in Section 4.1, now comprising all multinomial distributions on $\{0, 1, 2, \ldots, T + 1\}$ x $\{0, 1, 2, \ldots, T + 1\}$. The trial design draws $N(k)$ persons at random from J and assigns each sampled person to dose level $t_k$. For $j \in N(k)$, one observes $[d_j(t_k),\,e_j(t_k)]$. Thus, $\psi = \{N(k), p_{N(k)}[d(t_k),\,e(t_k)], k = 1, \ldots, K\}$.

Computation of finite-sample MMR decisions is typically a complex problem. I expect this to be so in the setting of this paper. I do not address the problem here, but it is an important topic for future research. A computationally simple approach to dosage choice with sample data is to estimate $p[d(t_k),\,e(t_k)]$, $k = 1, \ldots, K$ by their sample analogs $p_{N(k)}[d(t_k),\,e(t_k)]$, $k = 1, \ldots, K$ and then proceed as in Sections 4.1 and 4.2, acting as if the estimates are accurate. Manski (2021) discusses this "as-if" approach to decision making in abstraction and explains how to compute its maximum regret numerically.

An interesting question is to learn how the assumption of monotone dose response affects finite-sample inference and decision making, relative to what occurs with design-based trial analysis. Even when a complete dosing trial has been performed, enabling design-based analysis to point-identify dose response, the monotonicity assumption may still improve finite-sample analysis. It may improve the statistical precision of point estimates of outcome probabilities used in as-if decision making and thereby reduce the maximum regret achieved when estimates are used to choose dosage.

5. Discussion

I have studied dosage choice based on limited trial data. Medical decision making has relied on such data. Indeed, the Phase III evidence used in FDA drug approval is legally required to be trial data. Limited trial data are also the main source of evidence for other realms of decision making. A broad theme of this paper is that the standard design-based approach to analysis of trial data is unnecessarily restrictive in settings where credible assumptions make empirical findings informative across treatments. The assumption of monotone dose response is highly credible in dosage trials. Other assumptions relating outcomes across treatments may be credible in other contexts.

Observational data is also informative about dose response when assumptions ME and MT hold. Consider an observational study where $z_j \in (0, \ldots, T)$ denotes the dose received by patient j. The data for this patient are $[z_j, d_j(z_j), e_j(z_j)]$. Observation of a random sample of the population enables estimation of the population distribution $p[z, d(z), e(z)]$. For each $s \in (0, \ldots, T)$, Bayes Theorem gives $p[z = s, d(z), e(z)] = p[d(s), e(s)|z = s]p(z = s)$. Knowledge of $p(z = s)$ reveals nothing about $q(t_d, t_e)$. However, when $p(z = s) > 0$, knowledge of $p[d(s), e(s)|z = s]$ reveals that

(7a)   $p[d(s) = 0, e(s) = 0|z = s] \ = \ q(t_d \leq s, t_e > s|z = s).$

(7b)   $p[d(s) = 1, e(s) = 0|z = s] \ = \ q(t_d > s, t_e > s|z = s).$

(7c)   $p[d(s) = 0, e(s) = 1|z = s] \ = \ q(t_d \leq s, t_e \leq s|z = s).$

(7d)   $p[d(s) = 1, e(s) = 1|z = s] \ = \ q(t_d > s, t_e \leq s|z = s).$

It can be shown that these inequalities yield an easily computed informative bound on $q(t_d, t_e)$. However, it appears delicate to obtain a sharp bound. Details may be obtained from the author.

It would be valuable to extend the analysis of this paper in multiple ways. Section 4.3 mentioned study of finite-sample MMR decisions. Another direction for future research is to recognize that disease and AE outcomes may vary in severity rather than being binary (yes/no), as assumed here. When outcomes vary in severity, the distribution of outcomes may still be characterized by a threshold dose distribution $q\{t_d, t_e\}$, but now $t_d$ and $t_e$ is each a vector of thresholds for passage into increasingly high levels of severity.

Yet another direction is to recognize that the one-dimensional dosing decision studied here is often a component of a more general problem of choice of a regimen; for example, specification of the (dose, schedule, duration) for administration of a drug. The concept of monotone treatment extends to regimens.

One regimen is more intense than another if it combines a weakly larger dose, more frequent schedule, and longer duration. Analysis of regimens is complex when comparing alternatives that are more intense in some components but less intense in others.

A further complexity in evaluation of regimens is that the decision process may be dynamic rather than static as assumed in this paper. As successive doses are administered, one may observe disease-AE outcomes as time passes. This may enable updating of decisions as outcome data accumulate.